\documentclass{article}
\usepackage[latin9]{inputenc}
\usepackage{geometry}
\usepackage{float}
\usepackage{amsmath}
\usepackage{amssymb}
\usepackage{graphicx}
\usepackage[unicode=true,pdfusetitle,
 bookmarks=true,bookmarksnumbered=false,bookmarksopen=false,
 breaklinks=false,pdfborder={0 0 1},backref=section,colorlinks=false]
 {hyperref}
\usepackage{breakurl}

\makeatletter

\providecommand{\tabularnewline}{\\}

\usepackage{fullpage}\usepackage[justification=centering]{caption}\usepackage[all]{hypcap}\usepackage{microtype}

\makeatother

\begin{document}

\title{Asymptotics of the 3$j$ and 9$j$ Coefficients}

\author{Daniel Hertz-Kintish, Larry Zamick, Brian Kleszyk\\
 \textit{Department of Physics and Astronomy, Rutgers University,
Piscataway, New Jersey 08854}\\
 }

\date{\today}
\maketitle
\begin{abstract}
In this work we present the details of calculations we previously
performed for the large $j$ behavior of certain 3$j$ and 9$j$ symbols. 
\end{abstract}
In this paper we focus on equations (11 and 13), and (23 and 24) of
the work of Kleszyk and Zamick \cite{Kleszyk}. In particular we consider
the case when the total angular momentum $I$ is equal to $I_{\text{max}}-2n$
and $I_{\text{max}}\equiv4j-2$, and $n=0,1,2,...$ We take the limit
of large $j$ where $n$ becomes much smaller than $j$. For convenience,
we also define $J=2j$, where $j$ is the total angular momentum of
a single particle.

We first address the 3$j$ coefficient, using the formula Eq. (13)
of \cite{Kleszyk}, a derivation of which is contained in the work
of Racah \cite{Racah}:

\begin{equation}
\left(\begin{array}{ccc}
2j & 2j-2 & I\\
0 & 0 & 0
\end{array}\right)
\end{equation}
We express the total angular momentum $I$ using a new variable $m$
such that $I=4j-2m$, where this time $m=1,2,3,...$. We can separate
parts of the 3$j$ which now becomes 
\begin{equation}
3j=\dfrac{\sqrt{(2m-1)!}}{(m-1)!}(-1)^{m}\sqrt{\dfrac{N_{1}!N_{2}!}{N_{3}!}}\dfrac{N_{4}!}{N_{5}!N_{6}!}
\end{equation}
where the 6 factors $N_{i}$ are: 
\begin{equation}
\tag{2a}\begin{array}{ccc}
N_{1}=2J-2-2m & N_{2}=2J+2-2m & N_{3}=4J-1-2m\\
N_{4}=2J-1-m & N_{5}=J-1-m & N_{6}=J+1-m
\end{array}\label{eq:defN}
\end{equation}
We use the Stirling approximation, 
\begin{equation}
\ln x!=x\ln x-x+\ln\sqrt{2\pi x}
\end{equation}
and it should be noted that the approximation approaches the true
value asymptotically. Now we can write: $N_{i}=(\alpha_{i}+\beta_{i}m+\gamma_{i}J)$
with differing constant coefficients. In Eq.\eqref{eq:defN} we give
the contribuition of $-N,\ln\sqrt{2\pi N},\alpha\ln N,m\beta\ln N,$
and $\gamma J\ln N$. For the latter we break things up into (a) ``extreme''
and (b) ``next order''. This is necessary because ``next order''
has contributions comparable to those in ``$-N$''.

\begin{table}[h]
\centering \protect\protect\protect\protect\caption{Asymptotic contributions to the $3j$ coefficients}

\label{table:3j} %
\begin{tabular}{l|r|r|r|r|r|r}
 & $-N_{i}$  & $\ln\sqrt{2\pi N_{i}}$  & $\alpha_{i}\ln N_{i}$  & $\beta_{i}m\ln N_{i}$  & $\gamma_{i}J\ln N_{i}$  & $\gamma_{i}J\ln N_{i}$\tabularnewline
\hline 
(1)  & $\dfrac{}{}-\frac{1}{2}(2J-2-2m)$  & $\frac{1}{2}\ln\sqrt{4\pi J}$  & $-\ln(2J)$  & $\dfrac{}{}-m\ln(2J)$  & $J\ln2J$  & $-1-m$ \tabularnewline
(2)  & $\dfrac{}{}-\frac{1}{2}(2J+2-2m)$  & $-\frac{1}{2}\ln\sqrt{8\pi J}$  & $\ln(2J)$  & $\dfrac{}{}-m\ln(2J)$  & $J\ln(2J)$  & $1-m$ \tabularnewline
(3)  & $\dfrac{}{}\frac{1}{2}(4J-1-2m)$  & $\frac{1}{2}\ln\sqrt{4\pi J}$  & $\frac{1}{2}\ln(4J)$  & $\dfrac{}{}m\ln(4J)$  & $-2J\ln(4J)$  & $\frac{1}{2}+m$ \tabularnewline
(4)  & $\dfrac{}{}\frac{}{}-(2J-1-m)$  & $\frac{1}{2}\ln\sqrt{4\pi J}$  & $-\ln(2J)$  & $\dfrac{}{}-m\ln(2J)$  & $2J\ln(2J)$  & $-1-m$ \tabularnewline
(5)  & $\dfrac{}{}(J-1-m)$  & $-\frac{1}{2}\ln\sqrt{2\pi J}$  & $\ln(J)$  & $\dfrac{}{}m\ln J$  & $-J\ln J$  & $1+m$ \tabularnewline
(6)  & $\dfrac{}{}(J+1-m)$  & $-\frac{1}{2}\ln\sqrt{2\pi J}$  & $-\ln(J)$  & $\dfrac{}{}m\ln J$  & $-J\ln J$  & $-1+m$ \tabularnewline
\hline 
Total  & $\dfrac{}{}\frac{1}{2}$  & $\ln\left(\frac{2}{\pi J}\right)^{1/4}$  & $\ln\frac{1}{\sqrt{J}}$  & $\dfrac{}{}-m\ln2$  & $0$  & $-\frac{1}{2}$ \tabularnewline
\end{tabular}
\end{table}

First notice that ``$\gamma J\ln N$'' result is $\dfrac{1}{2}$,
which cancels the $+\dfrac{1}{2}$ from ``$-N$''. Adding up all
the totals we get 
\begin{equation}
-m\ln2+\ln\left(\dfrac{2}{\pi J}\right)^{1/4}+\ln\left(\dfrac{1}{\sqrt{J}}\right)
\end{equation}
\begin{equation}
=-m\ln2+\ln\left(\dfrac{2}{\pi J^{3}}\right)^{1/4}
\end{equation}
Taking the antilog we get 
\begin{equation}
3j\approx e^{m\ln2}\left(\dfrac{2}{\pi J^{3}}\right)^{1/4}
\end{equation}
and note that $e^{-m\ln2}=\dfrac{1}{2^{m}}$.

Putting everything together and putting things in terms of $j$ and
$n$ we obtain 
\begin{equation}
3j\rightarrow\dfrac{\sqrt{(2n)!}}{n!2^{n}}(-1)^{n}\left(\dfrac{1}{64\pi j^{3}}\right)^{1/4}
\end{equation}
We see that in the limit $n\ll j$, $3j$ goes as $\dfrac{1}{j^{3/4}}$.
Alternatively the Clebsch-Gordan has an asymptotic value 
\begin{equation}
CG\rightarrow\dfrac{\sqrt{(2n)!}}{n!2^{n}}(-1)^{n}\left(\dfrac{1}{\pi j}\right)^{1/4}
\end{equation}

We next consider the unitary $9j$ coefficient $\langle(jj)^{2j}(jj)^{2j}|(jj)^{2j}(jj)^{2j-2}\rangle^{I}$.
Again we will write $I=4j-2m$, with $m=1,2,3,...$. In Eq. (11) from
\cite{Kleszyk}, we have a factor $(2J+I+1)!$ which becomes $(4J+1-2m)!$.
This can be written as $(4J+1)!\times PROD$ where $PROD=(4J+1)(4J)...(4J+2-2m)$.
For convenience we break this equation into several parts: 
\begin{equation}
U(9j)=\dfrac{FAC}{\sqrt{PROD}}\sqrt{\dfrac{(2J+1)(2J-3)}{2}}\times3j
\end{equation}
where 
\begin{equation}
FAC=\dfrac{(C_{1}!)^{2}}{C_{2}!}\sqrt{\dfrac{C_{3}!}{C_{4}!C_{5}!}}
\end{equation}
with 
\[
\begin{array}{rrrrr}
C_{1}=J & C_{2}=2J & C_{3}=4J+1 & C_{4}=2J+1 & C_{5}=2J-1\end{array}
\]
There are $2m$ terms in $PROD$. We use the fact that $(4J+1-2m)!=(4J+1)!\times PROD$,
and asymptotically we obtain 
\begin{equation}
\sqrt{\dfrac{(2J+1)(2J-3)}{2}}\rightarrow J\sqrt{2}
\end{equation}
\begin{equation}
PROD\rightarrow(4J)^{2m}=(8j)^{2m}
\end{equation}
Hence we have 
\begin{equation}
\dfrac{1}{\sqrt{PROD}}\rightarrow\dfrac{1}{(8j)^{m}}
\end{equation}

We use the Stirling approximation to calculate $FAC$. The detailed
results are given in Table \ref{table:u9j}. 
\begin{table}[h]
\protect\protect\protect\protect\caption{ $\ln\dfrac{(C_{1}!)^{2}}{C_{2}!}\dfrac{C_{3}!}{C_{4}!C_{5}!}$ \label{table:u9j} }

\centering %
\begin{tabular}{r|r|r|r|r|r}
 & $-C_{i}$  & $\ln\sqrt{2\pi C_{i}}$  & $\alpha_{i}\ln C_{i}$  & $\gamma_{i}J\ln()$  & $\gamma_{i}J\ln()$ \tabularnewline
\hline 
(1)  & $\dfrac{}{}-2J$  & $2\ln(\sqrt{2\pi J})$  & $0$  & $2J\ln J$  & $0$ \tabularnewline
(2)  & $\dfrac{}{}+2J$  & $-\ln\sqrt{4\pi J}$  & $0$  & $-2J\ln(2J)$  & $0$ \tabularnewline
(3)  & $\dfrac{}{}-2J-\frac{1}{2}$  & $\ln\sqrt{8\pi J}$  & $\frac{1}{2}\ln(4J)$  & $2J\ln(4J)$  & $\frac{1}{2}$ \tabularnewline
(4)  & $\dfrac{}{}J+\frac{1}{2}$  & $-\frac{1}{2}\ln\sqrt{4\pi J}$  & $-\frac{1}{2}\ln(2J)$  & $-J\ln(2J)$  & $\frac{1}{2}$ \tabularnewline
(5)  & $\dfrac{}{}J-\frac{1}{2}$  & $-\frac{1}{2}\ln\sqrt{4\pi J}$  & $\frac{1}{2}\ln(2J)$  & $-J\ln(2J)$  & $-\frac{1}{2}$ \tabularnewline
\hline 
Total  & $\dfrac{}{}-\frac{1}{2}$  & $\ln(\frac{\pi J}{2})^{1/4}$  & $\ln(2\sqrt{J})$  & $0$  & $\frac{1}{2}$\tabularnewline
\end{tabular}
\end{table}

We next combine Tables 1 and 2. There are many cancellations when
we add the totals of $\ln FAC$ and $\ln3j$ in Table~\ref{table:3j}
and Table~\ref{table:u9j}. The result is 
\begin{equation}
\ln FAC+\ln3j=-(m-1)\ln2=-n\ln2
\end{equation}
The antilog is 
\begin{equation}
e^{-n\ln2}=\dfrac{1}{2^{n}}
\end{equation}
The $j$ dependence comes from 
\begin{equation}
\sqrt{\dfrac{(2J+1)(2J-3)}{2}}
\end{equation}
and PROD 
\begin{equation}
\sqrt{PROD}\rightarrow(8j)^{m}
\end{equation}
putting everything together we obtain the result 
\begin{equation}
U9j\rightarrow\dfrac{(-1)^{n}}{2\sqrt{2}16^{n}}\dfrac{\sqrt{\left((2n+2)!(2n)!\right)}}{(n!)j^{n}}.
\end{equation}
In the different limit of fixed $I$ and $j\gg I$, we get the behavior
\begin{equation}
U9j\rightarrow\sqrt{6\pi}j^{3/2}e^{-4\ln{2}j}.
\end{equation}

The best way to demonstrate the power-law behavior of the $U9j$ symbol
is to plot the logarithm of $U9j$ vs. the logarithm of $j$. We plot
this in Figure 1. Note the independence of the slopes of the curves
for different values of $n$.

\begin{figure}[h]
\centering \protect\protect\caption{$\ln|U9j|$ vs.$\ln j$ for many values of $n$}

\includegraphics[width=0.6\textwidth]{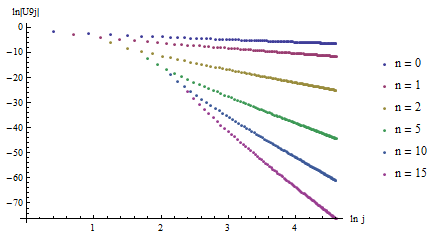} 
\end{figure}

We present results of the percent deviation of our approximate values
of $3j$ and $U9j$ from the exact values in Tables 3 and 4. 
\begin{table}[H]
\protect\protect\protect\caption{Comparison of the exact and asymptotic values of the 3j symbols}

\centering %
\begin{tabular}{crlll}
\hline 
 & j  & Accepted 3j  & Approximate 3j  & Percent Error\tabularnewline
\hline 
$n=0$  &  &  &  & \tabularnewline
 & 9/2  & 0.0917186951  & 0.0859524287  & 6.28690401\tabularnewline
 & 99/2  & 0.0143074760  & 0.0142302863  & 0.539505856\tabularnewline
 & 999/2  & 0.00251476295  & 0.00251342493  & 0.0532063347\tabularnewline
 & 9999/2  & 0.000446679154  & 0.000446655420  & 0.00531331215\tabularnewline
$n=1$  &  &  &  & \tabularnewline
 & 9/2  & -0.0703281160  & -0.0607775452  & 13.5800180\tabularnewline
 & 99/2  & -0.0101817625  & -0.0100623320  & 1.17298491\tabularnewline
 & 999/2  & -0.00177932008  & -0.00177725981  & 0.115789693\tabularnewline
 & 9999/2  & -0.000315869604  & -0.000315833077  & 0.0115641443\tabularnewline
$n=2$  &  &  &  & \tabularnewline
 & 9/2  & 0.0667864681  & 0.0526348981  & 21.1892774\tabularnewline
 & 99/2  & 0.00887471327  & 0.00871423511  & 1.80826305\tabularnewline
 & 999/2  & 0.00154190275  & 0.00153915215  & 0.178390316\tabularnewline
 & 9999/2  & 0.000273568204  & 0.000273519468  & 0.0178151485\tabularnewline
$n=10$  &  &  &  & \tabularnewline
 & 99/2  & 0.00642003383  & 0.00597328117  & 6.95872744\tabularnewline
 & 999/2  & 0.00106225244  & 0.00105503104  & 0.679819882\tabularnewline
 & 9999/2  & 0.000187614589  & 0.000187487331  & 0.0678293749\tabularnewline
$n=100$  &  &  &  & \tabularnewline
 & 999/2  & 0.000637437519  & 0.000596632653  & 6.40139073\tabularnewline
 & 9999/2  & 0.000106699870  & 0.000106026325  & 0.631251795\tabularnewline
\hline 
\end{tabular}\label{table:nonlin} 
\end{table}

\begin{table}[H]
\protect\protect\protect\caption{Comparison of the exact and asymptotic values of the U9j symbols}

\centering %
\begin{tabular}{crlll}
\hline 
 & j  & Accepted U9j  & Approximate U9j  & Percent Error\tabularnewline
\hline 
$n=0$  &  &  &  & \tabularnewline
 & 9/2  & 0.492152957  & 0.500000000  & 1.59443179\tabularnewline
 & 99/2  & 0.499361854  & 0.500000000  & 0.127792280\tabularnewline
 & 999/2  & 0.499937371  & 0.500000000  & 0.0125274006\tabularnewline
 & 9999/2  & 0.499993749  & 0.500000000  & 0.00125027349\tabularnewline
$n=1$  &  &  &  & \tabularnewline
 & 9/2  & -0.0378955625  & -0.0340206909  & 10.2251328\tabularnewline
 & 99/2  & -0.00312046463  & -0.00309279008  & 0.886872805\tabularnewline
 & 999/2  & -0.000306761485  & -0.000306492711  & 0.0876166329\tabularnewline
 & 9999/2  & -0.0000306243639  & -0.0000306216840  & 0.00875116429\tabularnewline
$n=2$  &  &  &  & \tabularnewline
 & 9/2  & 0.00606563844  & 0.00448261961  & 26.0981402\tabularnewline
 & 99/2  & 0.0000379552583  & 0.0000370464431  & 2.39443810\tabularnewline
 & 999/2  & $3.64686293*10^{-7}$  & $3.63819464*10^{-7}$  & 0.237691695\tabularnewline
 & 9999/2  & $3.63251097*10^{-9}$  & $3.63164818*10^{-9}$  & 0.0237519144\tabularnewline
$n=10$  &  &  &  & \tabularnewline
 & 99/2  & $7.33668833*10^{-17}$  & $5.24669432*10^{-17}$  & 28.4868855\tabularnewline
 & 999/2  & $4.95097802*10^{-27}$  & $4.79272848*10^{-27}$  & 3.19632873\tabularnewline
 & 9999/2  & $4.76517144*10^{-37}$  & $4.74976392*10^{-37}$  & 0.323335927\tabularnewline
$n=20$  &  &  &  & \tabularnewline
 & 99/2  & $1.75313503*10^{-27}$  & $5.21781167*10^{-28}$  & 70.2372517\tabularnewline
 & 999/2  & $4.88682624*10^{-48}$  & $4.35394087*10^{-48}$  & 10.9045287\tabularnewline
 & 9999/2  & $4.32566*10^{-68}$  & $4.27622870*10^{-68}$  & 1.143\tabularnewline
\hline 
\end{tabular}\label{table:nonlin} 
\end{table}

We note other work on asymptotics of CG coefficients by Reinsch and
Morehead \cite{Reinsch}. In their work they define

\begin{equation}
\beta=((j_{1}+j_{2}-j)(j+j_{2}-j_{1})(j+j_{1}-j_{2})(j_{1}+j_{2}+j))^{1/2}
\end{equation}
They find an approximate expression for the CG coeffecients in their
Eq.(B9).

\[
CG=\left<j_{1}j_{2}00|j0\right>\approx2(-1)^{\frac{j_{1}+j_{2}-j}{2}}\sqrt{\dfrac{2j+1}{2\pi\beta}}\sqrt{\dfrac{j+j_{1}+j_{2}}{j+j_{1}+j_{2}+1}}(1+\delta_{4}+\delta_{6})
\]

\begin{equation}
\times\left[1+\dfrac{1}{24}\left(\dfrac{2}{j}+\dfrac{2}{j_{1}}+\dfrac{1}{j_{2}}-\dfrac{1}{j+j_{1}+j_{2}}-\dfrac{1}{-j+j_{1}+j_{2}}-\dfrac{1}{j-j_{1}+j_{2}}-\dfrac{1}{j+j_{1}-j_{2}}\right)\right]
\end{equation}
We quickly run into trouble in making a comparison with our results,
especially for $n=0$. In their Eq.(B12) they have in the leading
term CG proportional to $\dfrac{1}{\sqrt{\beta}}$. However for the
case $j=j_{1}+j_{2}$, that is to say $I=I_{max}$, with our $n=0$,
we see that $\beta$ vanishes and hence their expression for CG blows
up. Evidently their formula is not valid in this region. On the other
hand, our expression Eq. (13) from \cite{Kleszyk} works just fine.

In this work, we have given the details of how the asymptotic behaviors
of selected $3j$ and $9j$ coefficients and their unitary counterparts
are obtained. There are some subtleties, e.g. in the second column
of Table 1, although term-by-term we get non-zero results, the entire
sum is zero and so we must expand further as in the following column.
There are similar points for Table 2. We further note that one can
take asymptotic limits in more than one way. Here the emphasis is
on when the total angular momentum $I$ is large ($I=I_{max}-2n,n\ll j$),
and one obtains a power-law behavior $1/j^{n}$ . This is most easily
seen by plotting $\ln|U9j|\text{ vs.}\ln j$. On the other hand, if
one keeps $I$ fixed and increases $j$ one gets a dominantly exponential
behavior, as shown in Eq. (19).This is most easily seen by plotting
$U9j$ vs. $j$. Lastly, we recall the physics motivation for this
work---how maximum-$j$ pairing manifests itself in nuclei \cite{Escuderos}.

Brian Kleszyk thanks the Rutgers Aresty Research Center for Undergraduates
for support during the 2013-2014 academic year. Daniel Hertz-Kintish
also thanks the Rutgers Aresty Research Center for Undergraduates
for support during the 2014 summer session.

\global\long\def\refname{

\end{document}
\vskip-1cm}
 
\begin{thebibliography}{1}
\bibitem[1]{Kleszyk} B. Kleszyk and L. Zamick, Analytical and Numerical
Calculations for the Asymptotic Behaviors of Unitary $9j$ Coefficients
Phys. Rev C.89.044322 (2014)

\bibitem[2]{Racah} G. Racah, Phys. Rev. 62, 438 (1942)

\bibitem[3]{Reinsch} M.W. Reinsch and J.J. Morehead, Journal of Mathematical
Physics 40, 4782 (1999)

\bibitem[4]{Varshalovich} D. A. Varshalovich, A. N. Moskalev, V.
K. Khersonskii, Quantum Theory of Angular Momentum, World Scientific,
Singapore (1988)

\bibitem[5]{Escuderos} L. Zamick and A. Escuderos, Phys. Rev. C.87.044302
(2013)\end{thebibliography}
